\documentclass[12pt]{article}
\usepackage[latin1]{inputenc}
\usepackage{cite}
\usepackage{amsmath}
\usepackage{amsfonts}
\usepackage{amssymb}
\usepackage{graphicx}
\usepackage{geometry}
\usepackage{amssymb,epsfig}
\usepackage{hyperref}

\makeatletter
\renewcommand\section{\@startsection {section}{1}{\z@}%
                                 {-3.5ex \@plus -1ex \@minus -.2ex}
                                   {2.3ex \@plus.2ex}%
                                   {\normalfont\large\bfseries}}
\renewcommand\subsection{\@startsection{subsection}{2}{\z@}%
                                   {-3.25ex\@plus -1ex \@minus -.2ex}%
                                     {1.5ex \@plus .2ex}%
                                     {\normalfont\bfseries}}
\renewcommand\subsubsection{\@startsection{subsubsection}{3}{\z@}%
                                   {-3.25ex\@plus -1ex \@minus -.2ex}%
                                     {1.5ex \@plus .2ex}%
                                     {\normalfont\itshape}}
\makeatother

\def\pplogo{\vbox{\kern-\headheight\kern -29pt
\halign{##&##\hfil\cr&{\ppnumber}\cr\rule{0pt}{2.5ex}&\ppdate\cr}}}
\makeatletter
\def\ps@firstpage{\ps@empty \def\@oddhead{\hss\pplogo}%
  \let\@evenhead\@oddhead 
}
\def\maketitle{\par
 \begingroup
 \def\thefootnote{\fnsymbol{footnote}}
 \def\@makefnmark{\hbox{$^{\@thefnmark}$\hss}}
 \if@twocolumn
 \twocolumn[\@maketitle]
 \else \newpage
 \global\@topnum\z@ \@maketitle \fi\thispagestyle{firstpage}\@thanks
 \endgroup
 \setcounter{footnote}{0}
 \let\maketitle\relax
 \let\@maketitle\relax
 \gdef\@thanks{}\gdef\@author{}\gdef\@title{}\let\thanks\relax}
\makeatother

\numberwithin{equation}{section}

\newcommand{\be}{\begin{equation}}
\newcommand{\bea}{\begin{eqnarray}}
\newcommand{\ee}{\end{equation}}
\newcommand{\eea}{\end{eqnarray}}

\newcommand{\Tr}{{\rm Tr}}

\begin{document}
 
\setcounter{page}0
\def\ppnumber{\vbox{\baselineskip14pt
}}
\def\ppdate{\footnotesize{}} \date{}

\author{Carlos Tamarit\\
[7mm]
{\normalsize  \it Perimeter Institute for Theoretical Physics}\\
{\normalsize \it Waterloo, ON, N2L 2Y5, Canada}\\
[3mm]
{\tt \footnotesize ctamarit at perimeterinstitute.ca}
}

\title{\bf Large, negative threshold contributions to light soft masses in models with Effective Supersymmetry
\vskip 0.5cm}
\maketitle

\begin{abstract} \normalsize
\noindent Threshold contributions to light scalar soft masses due to heavy sparticles (possibly including a heavy Higgs mostly aligned with $H_d$) in Effective SUSY scenarios are dominated by two-loop diagrams involving gauge couplings. This is due to the fact that in the limit in which the heavy states are degenerate, their one-loop contributions to the light soft masses only depend on small Yukawas and the hypercharge coupling. The two-loop threshold corrections involving only gauge couplings are calculated accounting for nonzero gaugino and light squark  masses and shown to be negative, and rather large ($\delta m^2_{t,L}\sim-480^2\,{\rm GeV}^2$ for heavy sparticles with masses around 10 TeV). The effect on tachyon bounds is revisited with calculations implementing decoupling. It is pointed out that models yielding Effective SUSY spectra using gaugino mediation require in general very heavy gluinos or a very low SUSY breaking scale in order to avoid tachyons (e.g. for heavy squarks at 10 TeV and a SUSY 
breaking 
scale of 
125 TeV, minimal scenarios require $\tilde m_3\gtrsim 2$ TeV at 500 GeV, while nonminimal ones demand  $\tilde m_3\gtrsim 8$ TeV). 
\end{abstract}
\bigskip
\newpage
\newpage

\section{Introduction}
Effective Supersymmetry (SUSY) scenarios \cite{Dimopoulos:1995mi,Cohen:1996vb,Brust:2011tb}, in which the first and second generation scalars of the Minimal Supersymmetric Standard Model (MSSM) are heavy --as well as possibly  some of the third generation scalars, always excluding the left-handed quark doublet and the right-handed stop-- remain  well-motivated realizations of Supersymmetry  which are natural, solve the flavor problem and are poorly constrained by the ongoing searches at the LHC due to the difficulty in separating light stop signals from top quark backgrounds \cite{Plehn:2012pr,Han:2012fw}.

In the absence of strong experimental constraints, some theoretical ones have been known for a while. It was pointed out in ref.~\cite{ArkaniHamed:1997ab} that, in the case of high-scale SUSY breaking, the Renormalization Group (RG) effects of the heavy scalars can drive the soft masses of the light third generation scalars towards tachyonic values, opening the possibility of phenomenologically disfavored charged or colored vacua. In ref.~\cite{Tamarit:2012ie} it was pointed out that the large hierarchies in the spectrum of sparticles called for an analysis that explicitly implemented the decoupling of heavy particles, which was shown to relax the tachyon bounds coming from the study of the RG evolution \cite{Tamarit:2012ry}.

In keeping with the idea of performing accurate calculations in Effective SUSY scenarios, it is necessary to examine the effect of finite threshold contributions due to the heavy sparticles at the scale at which they are integrated out. The threshold contributions to the soft masses of the light  scalars will involve, on dimensional grounds, the masses of the heavy particles in the loops, and thus are expected to be significant. One-loop threshold effects in the MSSM are well known \cite{Pierce:1996zz}. It turns out that in the limit of degenerate heavy scalars --and, if a heavy Higgs state is present, in scenarios in which it is mostly aligned with $H_d$-- these one-loop contributions only depend on small Yukawa couplings and the hypercharge gauge coupling, and may be negative. Results for two-loop threshold corrections due to heavy fields have been obtained in refs. \cite{Agashe:1998zz} and \cite{Hisano:2000wy}, following the results in ref. \cite{Martin:1996zb}, but neglecting the masses of the light 
sparticles. 

Since two-loop diagrams involve the strong gauge coupling, the previous observations suggest that they can be the dominant contributions to the threshold corrections of the light soft masses, or could be relevant to compensate for the one-loop tachyonic contributions. As previous computations ignored the soft masses of gluinos and the light squarks, this paper presents the corresponding results when they are taken into account. This is the proper thing to do when performing the computations by integrating heavy particles at their thresholds: first, the hidden sector fields that break SUSY  are integrated out, yielding the MSSM with nonzero soft masses, and next the heavy MSSM scalars are also integrated out at their corresponding scales. The calculations for the threshold contributions of the heavy scalars are performed in the $\overline{\rm MS}$ scheme, and they can be directly matched with the results for the low energy, nonsupersymmetric theories obtained after 
decoupling the heavy states in refs.~\cite{Tamarit:2012ie,Tamarit:2012ry}. The computation is similar in spirit to that leading to the scalar soft masses in gauge mediation; differences stem from the absence of loops of massive fermions, the absence of mixing of the heavy scalars, the presence of new hypercharge dependent contributions, and the fact that nonzero masses for the gluinos and the light  scalars are considered in the propagators.

The result is that these two-loop corrections evaluated at the threshold of the heavy sparticles are negative, and rather large; also, nonzero gluino masses have a sizable impact and tend to enhance the threshold effects, while the dependence on the masses of the light scalars is weaker. This of course contradicts the analogy with minimal gauge mediation, which may have suggested that the first and second generation fields could act as messenger fields that transmit SUSY breaking to the third generation; rather, the heavy fields tend to destabilize the light scalars.  

These negative threshold effects call for a reappraisal of the lower bounds for the high scale boundary values of the light scalar masses obtained by demanding the absence of tachyonic squarks and sleptons. Also, in models in which the third generation squark masses arise as a result of gaugino mediation \cite{Craig:2011yk,Craig:2012hc,Cohen:2012rm}, one may obtain in a similar way lower bounds for gaugino masses, since these will have to be large enough to compensate for the tachyonic RG and finite corrections. 

The paper is organized as follows. One-loop gauge-coupling dependent contributions are reviewed in section~\ref{sec:oneloop}. Section~\ref{sec:twoloop} centers on the two-loop contributions. In view of the results, tachyon bounds on high scale light scalar masses are revisited in section~\ref{sec:bounds} using the two-loop RG equations of ref.~\cite{Tamarit:2012ie} supplemented with the threshold corrections obtained in this paper; similarly, bounds on gaugino masses are obtained in  models involving gaugino mediation for the third generation. Section~\ref{sec:models} summarizes the results.

\section{\label{sec:oneloop}One loop contributions}
Neglecting off-diagonal Yukawas and a-terms that mix light and heavy scalars in Effective SUSY scenarios, and assuming degenerate heavy states with mass $M$,
the one-loop threshold correction at a scale $\mu$ in the $\overline{\rm MS}$ scheme in the Feynman gauge for a light soft mass $m^2_i$  due to the heavy squarks and sleptons  is
\begin{equation}
 \label{eq:oneloopthresh}
 \delta (m^2_i)^{1\,\rm loop}_{\tilde q,\tilde l}(\mu)=-\frac{g_1^2}{16\pi^2}Y_i\sum_j({d}_j Y_j)M^2\left(1-\log\frac{M^2}{\mu^2}\right).
\end{equation}
This contribution, which comes from diagrams with a quartic vertices coming from D-terms, is included in the general formulae of ref.~\cite{Pierce:1996zz}, which are written for the nondegenerate case and include nonzero mixing angles. In the expression above, $Y$ designates hypercharges. The sum in $j$ is over all the U(1) representations of the heavy scalars fields, whose dimension is denoted by ${d}_j$. In minimal Effective SUSY scenarios, the heavy scalars include those of the first two generations plus the sleptons and right-handed sbottom of the third generation, yielding  $\sum_j({d}_j Y_j)=1$, while in nonminimal scenarios all fields in the third generation are light, which gives $\sum_j({d}_j Y_j)=0$. The absence of contributions dependent on the gauge couplings $g_2,\,g_3$ is due to the degeneracy of the heavy fields and the identities $\Tr\,T^a=0$ for SU(2) and SU(3) groups. 

In Effective SUSY models in which the combination of Higgs doublets 
\begin{align}
{\cal H}_{\rm heavy}=\sin\alpha H_u-\cos\alpha H^\dagger_d
\label{eq:heavyH}
\end{align}
is also made heavy as well, and assuming that it also has a mass $M$, the formula above is still valid if the following substitution is used
\begin{align}\nonumber
  &\sum_j({d}_j Y_j)=2\sin^2\alpha,\quad {\text{ minimal Effective SUSY scenarios with a single light Higgs}},\\
   &\sum_j({d}_j Y_j)=-\cos2\alpha,\quad {\text{ nonminimal Effective SUSY scenarios with a single light Higgs}}\label{eq:sumY}.
\end{align}
The heavy Higgs has additional contributions proportional to diagonal Yukawas and a-terms. The former are again quadratic in the heavy mass $M$, while the latter are proportional to the trilinear couplings squared, and may be neglected assuming $a_i\ll M$. In this way one obtains the following threshold contributions due to the heavy Higgs field (ignoring again off-diagonal Yukawas):
\begin{align}
 \label{eq:oneloopthreshH}\delta (m^2_Q)^{1\,\rm loop}_{\cal H}(\mu)&=-\frac{1}{16\pi^2}(y_t^2\sin^2\alpha+y_b^2\cos^2\alpha)M^2\left(1-\log\frac{M^2}{\mu^2}\right),\\
 \nonumber\delta (m^2_U)^{1\,\rm loop}_{\cal H}(\mu)&=-\frac{1}{8\pi^2}y_t^2\sin^2\alpha M^2\left(1-\log\frac{M^2}{\mu^2}\right),\\
  \nonumber\frac{1}{y_b^2}\,\delta (m^2_D)^{1\,\rm loop}_{\cal H}(\mu)&=\frac{2}{y_\tau^2}\,\delta (m^2_L)^{1\,\rm loop}_{\cal H}(\mu)=\frac{1}{y_\tau^2}\,\delta (m^2_E)^{1\,\rm loop}_{\cal H}(\mu)=-\frac{1}{8\pi^2}\cos^2\alpha M^2\left(1-\log\frac{M^2}{\mu^2}\right).
\end{align}
The Yukawa couplings in the formulae above are those in the MSSM. Clearly, the threshold contributions of eqs.~\eqref{eq:oneloopthresh} and \eqref{eq:oneloopthreshH} may be sizable and negative when evaluated at the scale $\mu=M$ at which the heavy particles are integrated out. The Higgs contributions can be made small by choosing small values of $\alpha$, since then $\cal H$ is in the direction of $H_d$ and couples through the small Yukawas $y_b,y_\tau$. (However, in a consistent Higgs decoupling limit one has $\alpha\sim \frac{\pi}{2}-\beta$, where $\tan\beta=\frac{v_u}{v_d}$ is the ratio of Higgs VEVs \cite{Tamarit:2012ry}. Therefore small $\alpha$ implies large $\tan\beta$, which enhances the down Yukawas, so that $\alpha$ should not be too small). As an example, for $\cot\alpha\sim\tan\beta=10$  in nonminimal scenarios with a single Higgs, $M=\mu=10 {\rm TeV}$, one has the following threshold contributions, obtained by using the RG equations of ref.~\cite{Tamarit:2012ie} and matching the couplings with 
the experimental data as in ref.~\cite{Tamarit:2012ry}:
\begin{align*}
 \nonumber&\delta m^2_{\tilde q_L}\sim-93^2 {\rm GeV}^2,\quad \delta m^2_{\tilde t_R}\sim -254^2 {\rm GeV}^2, \quad \delta m^2_{\tilde b_R}\sim -96^2 {\rm GeV}^2,\quad \delta m^2_{\tilde L}\sim -222^2 {\rm GeV}^2,\\
&\delta m^2_{\tilde e_R}\sim 265^2 {\rm GeV}^2,\quad \delta m^2_{H_u}\sim -222^2 {\rm GeV}^2.
\end{align*}

\section{\label{sec:twoloop}Gauge-coupling dependent two loop diagrams involving heavy scalars}
\begin{figure}[t]\centering
\includegraphics{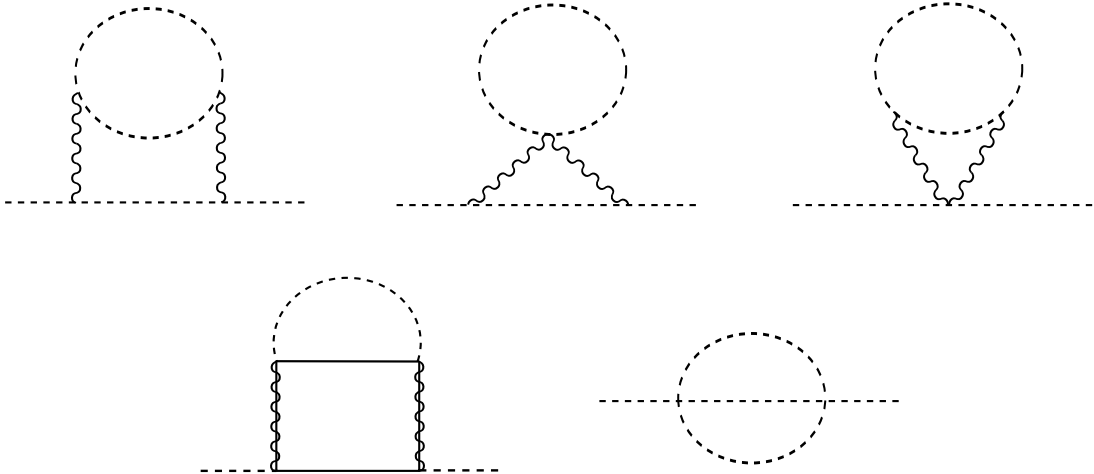}
\caption{\label{fig:2loopd} Two-loop diagrams involving heavy scalars contributing to the soft masses of the light scalars and not involving traces over hypercharges.}
\end{figure}

\begin{figure}[t]\centering
\includegraphics{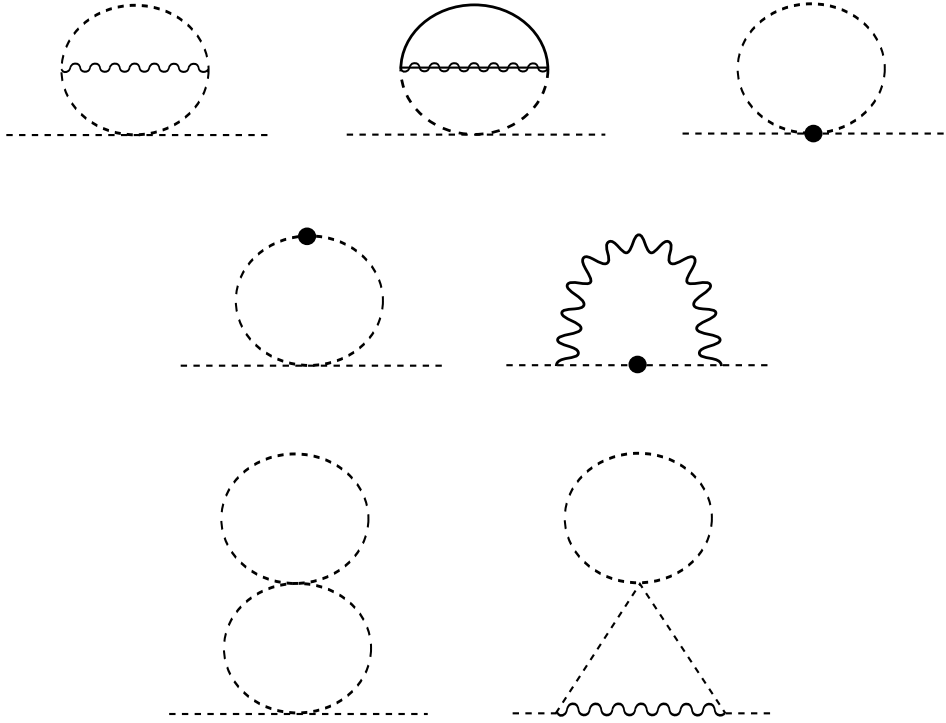}
\caption{\label{fig:2loopd2} Two-loop diagrams involving heavy scalars contributing to the soft masses of the light scalars and involving traces over hypercharges. The black dots represent one-loop counterterms}
\end{figure}

The results of the previous section show that one-loop finite corrections to the light soft masses due to degenerate heavy scalars may be negative when evaluated at the  corresponding thresholds. Since for small $\alpha$ they only involve small Yukawas and the hypercharge coupling,  this brings up the question of whether two-loop diagrams, which will also feature the strong gauge coupling, may or not partially cancel them. In the small $\alpha$ limit in which heavy states couple through small Yukawas --or when there is no heavy Higgs field-- 2 loop diagrams involving these couplings will be suppressed (more so than at one-loop level, since higher powers of the Yukawas will be present in general). Also, assuming $a_i\ll M$, diagrams with trilinear scalar couplings will be subdominant. Hence, in these scenarios the two-loop diagrams depending on gauge couplings are expected to be dominant. The diagrams that have nonzero, gauge-coupling dependent contributions to the soft masses of the light sparticles at two 
loops and don't involve traces over hypercharge are shown schematically in figure~\ref{fig:2loopd}. They are similar to the diagrams with internal scalar lines that yield soft masses for the MSSM scalars in minimal gauge mediation; in this case the propagators corresponding to messenger scalars are substituted by lines of heavy squarks, sleptons or Higgs fields. In contrast with the case of gauge mediation, the diagrams featuring traces over hypercharges do not necessarily add up to zero and therefore have to be included; they are shown in figure~\ref{fig:2loopd2}.

Assigning a nonzero mass $m_i$ to the $i$th light scalar and a mass $\tilde m_k$ to the gaugino of the $k$th group, the result of the diagrams in figure \ref{fig:2loopd} when all mixing angles between scalars are zero is, after proper subtraction in the $\overline{\rm MS}$ scheme in the Feynman gauge,
\begin{align}\nonumber
 &(\delta m^2_i)^{2\,\rm loop}(\mu)=-\frac{1}{3072\pi^4}\sum_{k,j}g^4_k C^{(k)}_iS^{(k)}_j\left\{M^2\left(16\pi^2-48-96\log\frac{M^2}{\mu^2}+24\phi\left[\frac{m^2_i}{4M^2}\right]\right)\right.\\
\nonumber &-m^2_i\left(42+\pi^2\!-36\log\frac{M^2}{\mu^2}\!+12\log^2\frac{M^2}{\mu^2}\!-36\log\frac{m^2_i}{M^2}+24\log\frac{M^2}{\mu^2}\log\frac{m^2_i}{M^2}\right.\\
 \nonumber&\left.\left.+6\log^2\frac{m^2_i}{M^2}+6\phi\left[\frac{m^2_i}{4M^2}\right]\right)-\frac{12}{{\tilde m}^2_k}\Big({\tilde m}^2_k \left(-4 M^2+{\tilde m}^2_k \left(-18+\pi ^2\right)\right)+\left(4 M^4+8 M^2 {\tilde m}^2_k\right.\right.\\
 \nonumber&\left.-6 {\tilde m}^4_k\right) \log^2\frac{M^2}{\mu^2}+4 \left({\tilde m}^2_k (M^2+5 {\tilde m}^2_k)+(M^2-{\tilde m}^2_k) (M^2+3 {\tilde m}^2_k) \log\frac{M^2-{\tilde m}^2_k}{\mu^2}\right) \log\frac{{\tilde m}^2_k}{\mu^2}\\
  \nonumber&-6 {\tilde m}^4_k \log^2\frac{{\tilde m}^2_k}{\mu^2}-4 \log\frac{M^2}{\mu^2} \Big(M^2 {\tilde m}^2_k+(M^2-{\tilde m}^2_k) (M^2+3 {\tilde m}^2_k) \log\frac{M^2-{\tilde m}^2_k}{\mu^2}+M^2 (M^2\\
 &\left.\left.+2 {\tilde m}^2_k) \log\frac{{\tilde m}^2_k}{\mu^2}\Big)+4 (M^2-{\tilde m}^2_k) (M^2+3 {\tilde m}^2_k) \text{Li}_2\left[\frac{{\tilde m}^2_k}{M^2}\right]\right)\right\}.
 \label{eq:2loopmassive}
\end{align}
In the previous formula, $\text{Li}_2$ is the dilogarithm function and $\phi$ is defined in eq.~\eqref{eq:phidef}. $C_i^{(k)}=\sum_a T_i^{k,a}T_i^{k,a}$ represents the Casimir of the gauge group $k$ in the representation $i$. For each value of $k$, the sum in $j$ runs over the irreducible representations (irreps) of the heavy scalars with respect to the $k$th gauge group, and  $S^{(k)}_j=\Tr \,T_j^{k,a}T_j^{k,a}$ (no sum over repeated indices) is the Dynkin index of the irrep labeled by $j$. $M$  designates again the mass of the heavy scalars; the result when these are nondegenerate can be simply obtained by substituting $M$ with $M_j$, allowing for different masses for the different representations of the heavy fields. In order to compute these mass corrections, the external momenta were set to zero from the beginning; the integrals were obtained in dimensional regularization using the formulae and techniques of refs.~\cite{Smirnov:2006ry} and \cite{Davydychev:1992mt}. More details are given in appendix \
ref{app:integrals}
. If the light scalars and the gauginos in the loops are massless ($m_i={\tilde m}_k=0$), the result is
\begin{align}
(\delta {m^2_i})^{2\,\rm loop,\,m_i=0}(\mu)=-\frac{M^2}{192\pi^4}\sum_{k,j}g^4_k C^{(k)}_iS^{(k)}_j\left(\pi^2-3-6\log\frac{M^2}{\mu^2}\right).
\label{eq:2loopmassless}
\end{align}
The last two terms inside the brackets differ from the corresponding result of ref.~\cite{Hisano:2000wy}, which was obtained from the formulae for soft masses in models of gauge mediation with generalized messenger sectors  by taking the limit in which the fermions in the loop become massless --in this paper, the diagrams that do not involve internal lines of heavy scalars were altogether ignored. The difference can be traced back to a different regularization of the infrared divergences: the authors of ref.~\cite{Hisano:2000wy} use an explicit infrared mass $m^2_\epsilon$ in the integrals denoted as $I[m_1,m_2,1,1,2]$ in appendix \ref{app:integrals} of this paper --see eq.~\eqref{eq:masterI}-- while the calculations presented here simply use dimensional regularization without additional regulators\footnote{While the use of $m^2_\epsilon$ is useful to separate UV and IR divergences and check the cancellation of the latter in physical observables, the use of dimensional regularization alone is equally valid 
for computing the 
same observables; however, due to the 
different regulators, the finite parts can differ, as happens in this case.}.

The contributions of the diagrams in figure 2  involving heavy scalars are as follows:
\begin{align}
  \nonumber&(\delta m^2_i)^{2\,\rm loop}(\mu)_Y=\sum_{j,k}g^2_1 g^2_k C^{(k)}_j {d_j} Y_jY_i\frac{M^2}{384\pi^4}\left(\!-9\!+\!\pi^2\!+\!6\log\frac{M^2}{\mu^2}\right)\!+\!\sum_{j,\hat i}g_1^4{d}_j{d}_{\hat i} Y^2_{j} Y_{\hat i} Y_i\frac{m^2_{\hat i}}{1536\pi^4}\times\\
  \nonumber&\left(6+\pi^2+3\log^2\frac{m_{\hat i}^2}{\mu^2}+6\log\frac{m_{\hat i}^2}{\mu^2}\left(-1+\log\frac{M^2}{\mu^2}\right)+3\left(-2+\log\frac{M^2}{\mu^2}\right)\log\frac{M^2}{\mu^2}\right)\\
  \nonumber&+\sum_{j,\hat j}g_1^4{d}_j{d}_{\hat j} Y^2_{\hat j} Y_j Y_i\frac{M^2}{256\pi^4}\log\frac{m_{\hat j}^2}{\mu^2}\left(-1+\log\frac{M^2}{\mu^2}\right)-\sum_{k, j}g^2_k g_1^2 d_{ j}C^{(k)}_{ j}Y_i Y_{j}\tilde m^2_k\frac{1}{768\pi^4}\Big(\pi^2\\
  &+6\log^2\frac{M^2}{\mu^2}\Big)
  \label{eq:2loophypercharge}.
\end{align}
The sum over $j$ runs over the irreducible representation of the heavy scalars; the sum over $\hat i$ is over those of the light scalars, while the sum in $\hat j$ is taken over both light and heavy fields (however, when evaluating the threshold corrections at the scale $\mu=M$, only the light fields will contribute to the sum in $\hat j$). $k$ runs over the Standard Model gauge groups, and ${d}_j$ denotes the dimension of the representation $j$. Regarding counterterms and their insertion, the $\overline{\rm MS}$ scheme was implemented by redefining $\mu$ as $\mu\rightarrow e^\gamma(4\pi)^{-1}\mu$ and then performing minimal subtraction. The contributions in eq.~\eqref{eq:2loophypercharge} proportional to $M^2$ coincide with the corresponding results in ref.~\cite{Hisano:2000wy}.

The dominant contributions are those proportional to the prefactor  $p^i\equiv\sum_{k,j}g^4_k C^{(k)}_iS^{(k)}_j$, since it includes terms that depend on the strong gauge coupling. In minimal and nonminimal Effective SUSY scenarios (denoted by MES and NMES) one has, respectively --this time neglecting the mixing angles of the heavy Higgs state, assuming it is mostly aligned with $H_d$:\footnote{Considering nonzero mixing angles between $H_u$ and $H_d$ modifies some diagrams, which become equivalent to vacuum integrals with three different masses in the propagators; these can be obtained from ref.~\cite{Davydychev:1992mt}. Since we are interested in the small $\alpha$ limit and since the heavy Higgs contributions  are subdominant with respect to those from fields charged under SU(3), we will not provide the full expressions.}
$$p^i_{\rm MES}=9g_1^4C^{(1)}_i+5g_2^4C^{(2)}_i+\frac{9}{2}g_3^4C^{(3)}_i,\quad p^i_{\rm NMES}=\frac{43}{6}g_1^4C^{(1)}_i+\frac{9}{2}g_2^4C^{(2)}_i+4g_3^4C^{(3)}_i,$$

\begin{figure}[t]\centering
\begin{minipage}{0.5\textwidth}
\includegraphics[width=8cm]{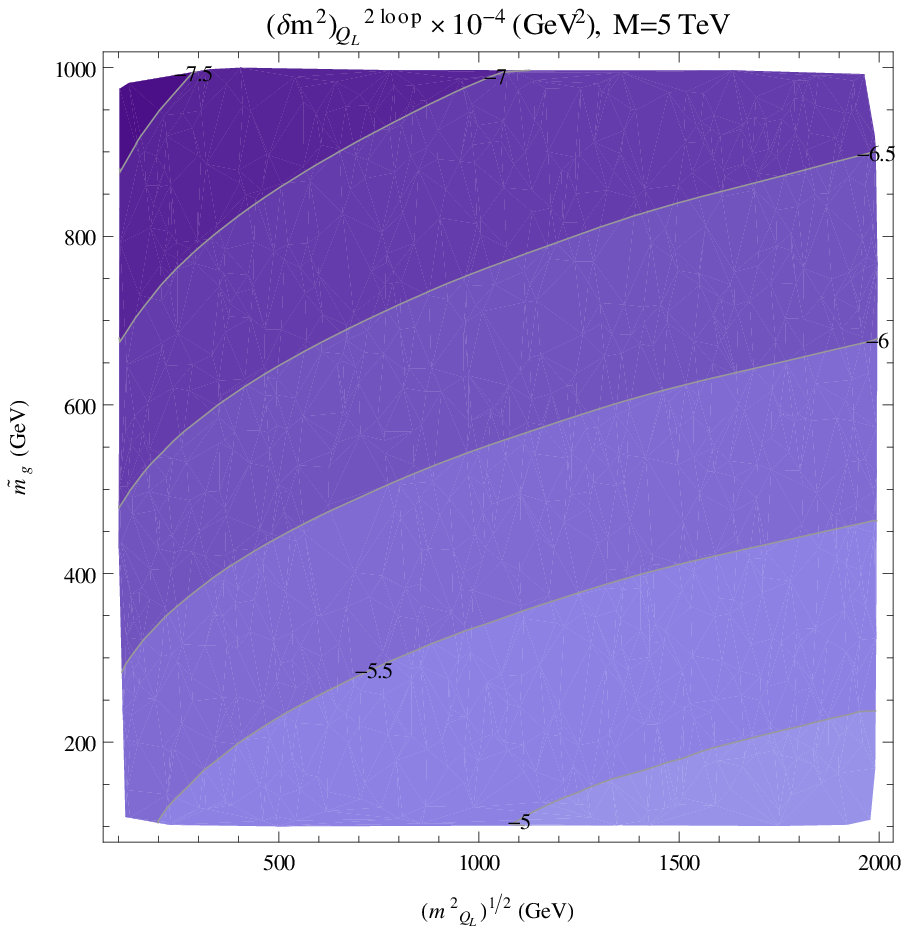}
\end{minipage}%
\begin{minipage}{0.5\textwidth}
\includegraphics[width=8cm]{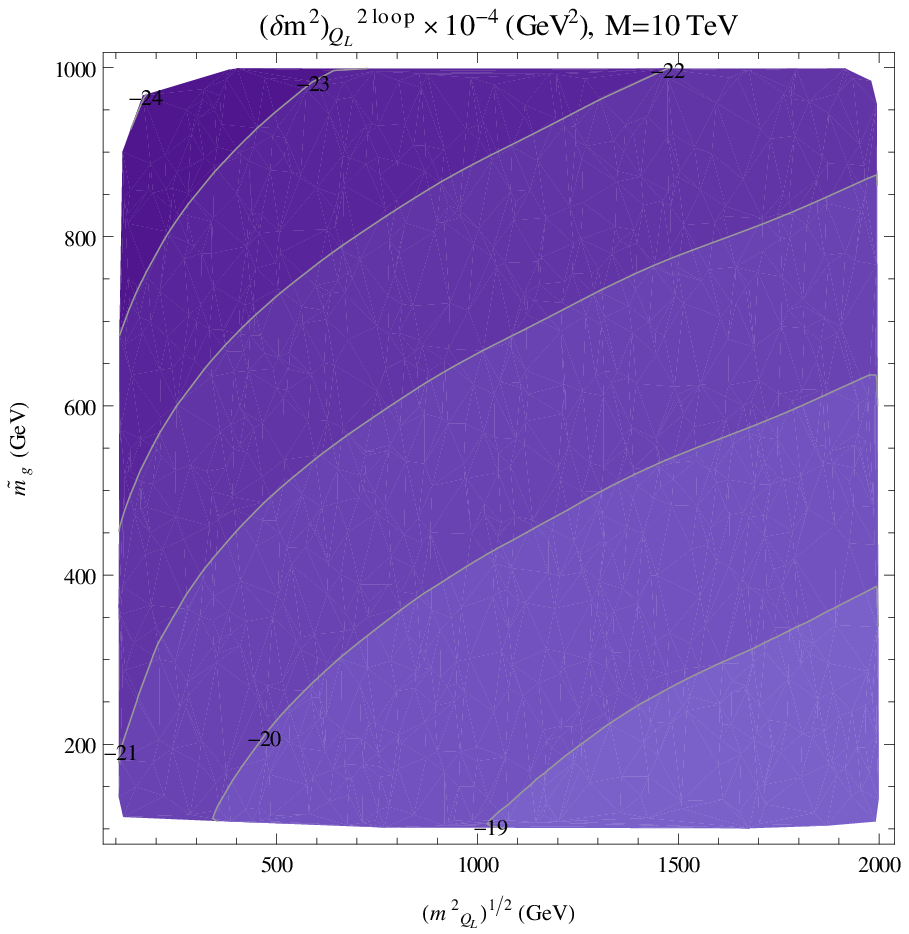}
\end{minipage}
\caption{\label{fig:2loopthresh} Two-loop threshold corrections to $m^2_{\tilde q_L}$ in terms of its tree-level value and a common gaugino mass $\tilde m_g$, for heavy sparticles at 5 TeV (left) and 10 TeV (right)}
\end{figure}

To recover the example from the end of the previous section, fixing $M=\mu=10\,{\rm TeV}$, $m_i=\tilde m_k=300\,{\rm GeV}$, $\cot\alpha=10$ in a nonminimal scenario, using eqs.~\eqref{eq:2loopmassive} and \eqref{eq:2loophypercharge} one gets
\begin{align*}
 \nonumber&\delta m^2_{\tilde q_L}\sim-483^2 {\rm GeV}^2,\quad \delta m^2_{\tilde t_R}\sim -462^2 {\rm GeV}^2, \quad \delta m^2_{\tilde b_R}\sim -460^2 {\rm GeV}^2,\quad \delta m^2_{\tilde L}\sim -151^2 {\rm GeV}^2,\\
&\delta m^2_{\tilde e_R}\sim -71^2 {\rm GeV}^2,\quad \delta m^2_{\tilde H}\sim -151^2 {\rm GeV}^2,
\end{align*}
where $m^2_{\tilde H}$ is the soft mass of the light Higgs field. Again, these numbers were obtained after computing the gauge couplings at the threshold scale as in ref.~\cite{Tamarit:2012ry}. Fig.~\ref{fig:2loopthresh} shows values of the 2 loop threshold corrections for $m^2_{\tilde q_L}$ as a function of the tree-level scalar mass and a common mass $\tilde m_g$ for the gauginos, for two different values of the heavy mass $M$. It is apparent that 2 loop corrections can be quite large and dominate over the one-loop contributions; also, heavy gauginos tend to enhance them.

\section{\label{sec:bounds}Tachyon bounds for squarks and gaugino masses}
It is known that the 2-loop renormalization group flow in the MSSM when the first and second generation sparticles are heavy may drive the light soft masses towards negative values, which would endanger the stability of the electroweak vacuum. Demanding the absence of tachyonic values for soft masses other than those of the Higgs allows to set lower bounds on the mass scales that set the boundary conditions for the RG flow at the SUSY breaking scale, which can be correlated with a lower bound on the amount of fine-tuning of the theory. As stated in the introduction, bounds were first calculated in ref.~\cite{ArkaniHamed:1997ab} using the RG MSSM equations in the $\overline{\rm DR}$ scheme; they were revisited in ref.~\cite{Tamarit:2012ry} after it was pointed out \cite{Tamarit:2012ie} that mass-independent schemes such as $\overline{\rm DR}$ or $\overline{\rm MS}$, being unphysical and not sensitive to mass thresholds, lack precision when large hierarchies in the masses are present, as in Effective SUSY 
scenarios. Using RG equations implementing 
decoupling, which effectively resum some of the large perturbative corrections, the tachyon bounds were shown to be substantially relaxed. 

\begin{figure}[h]\centering
 \begin{minipage}{.5\textwidth}\centering
   \includegraphics[scale=.9]{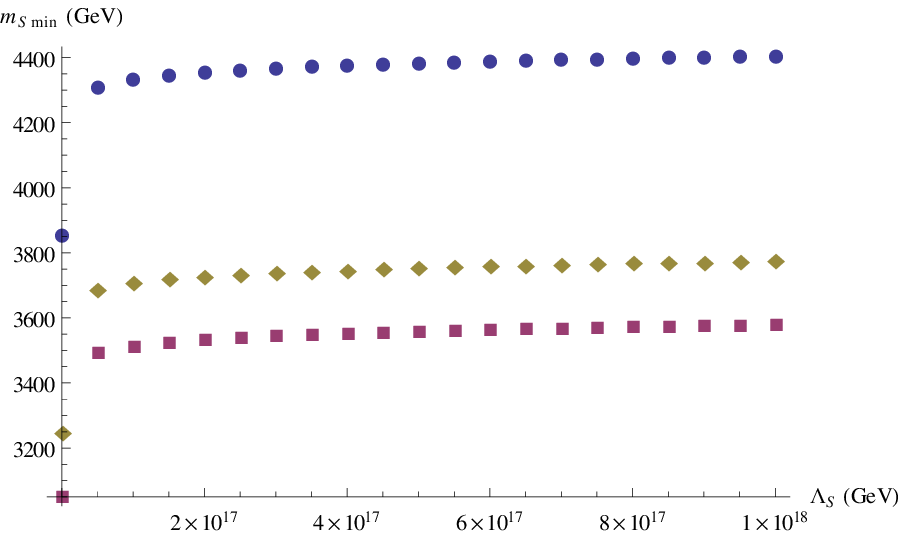}
\end{minipage}\begin{minipage}{0.5\textwidth}\centering
 \includegraphics[scale=.9]{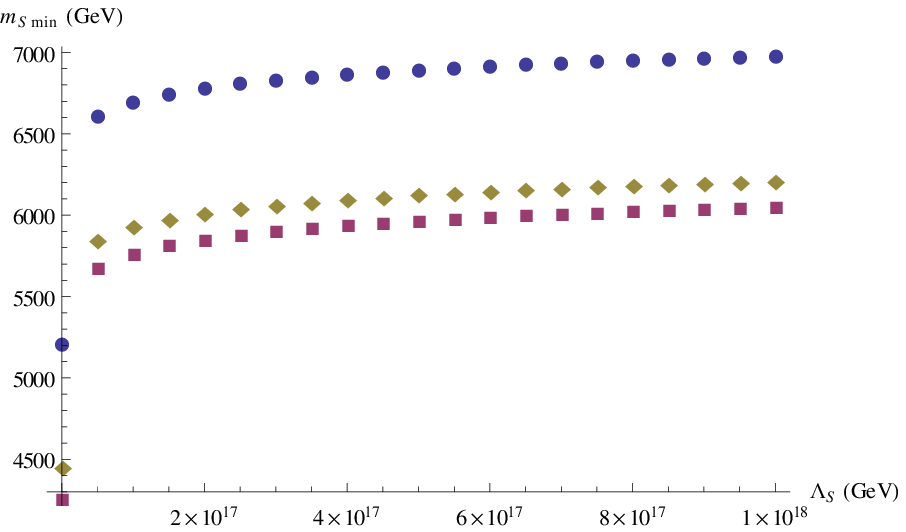}
\end{minipage}
\caption{\label{fig:tachyonmsugra} Minimum value of the scalar mass $m_s$ needed to avoid tachyonic soft masses at 500 GeV in terms of the high scale $\Lambda_S$, in minimal (left) and nonminimal (right) Effective SUSY scenarios with the boundary conditions of eq.~\eqref{eq:msugrabc}, with heavy sparticles at  20 TeV. The upper blue dots correspond to the the MSSM  $ \overline{\rm DR}$ RG flow, the diamond-shaped marks represent the results with the flow implementing  decoupling and including threshold effects, while the boxes denote the results when using the flow implementing decoupling but ignoring threshold effects.}
\end{figure}

\begin{figure}[h]\centering
 \begin{minipage}{.5\textwidth}\centering
   \includegraphics[scale=.9]{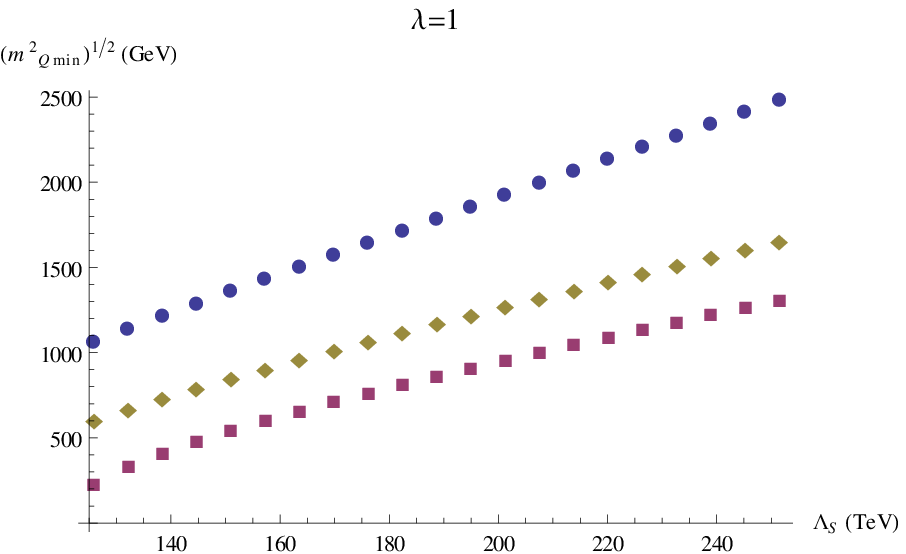}
\end{minipage}\begin{minipage}{0.5\textwidth}\centering
 \includegraphics[scale=.9]{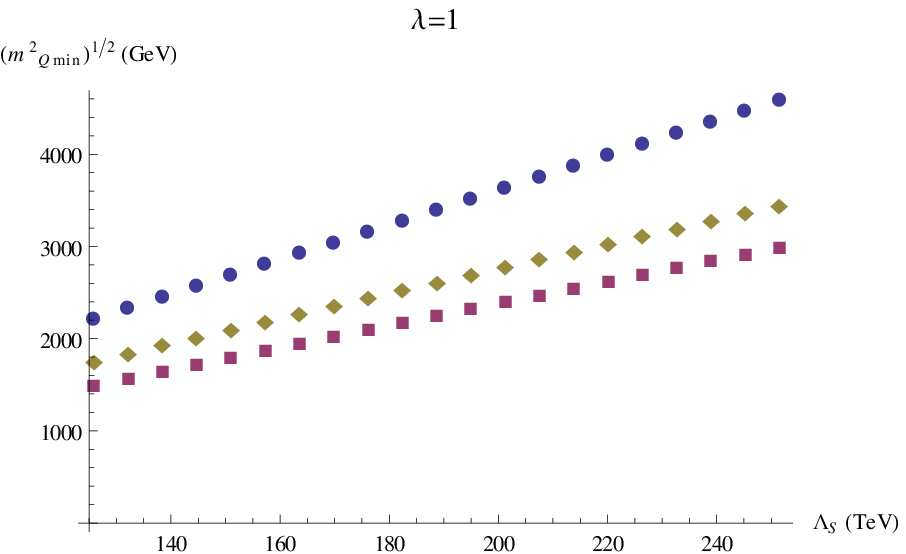}
\end{minipage}\vskip3mm
 \begin{minipage}{.5\textwidth}\centering
   \includegraphics[scale=.9]{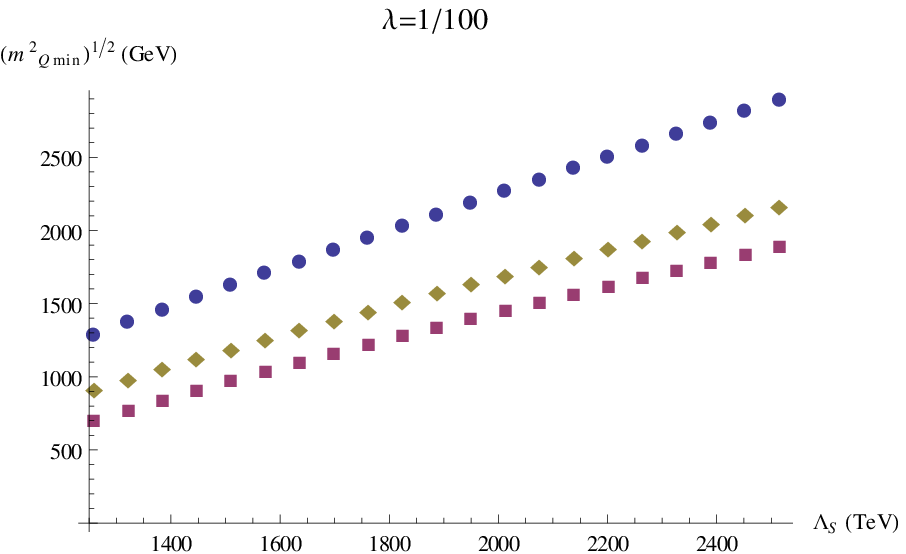}
\end{minipage}\begin{minipage}{0.5\textwidth}\centering
 \includegraphics[scale=.9]{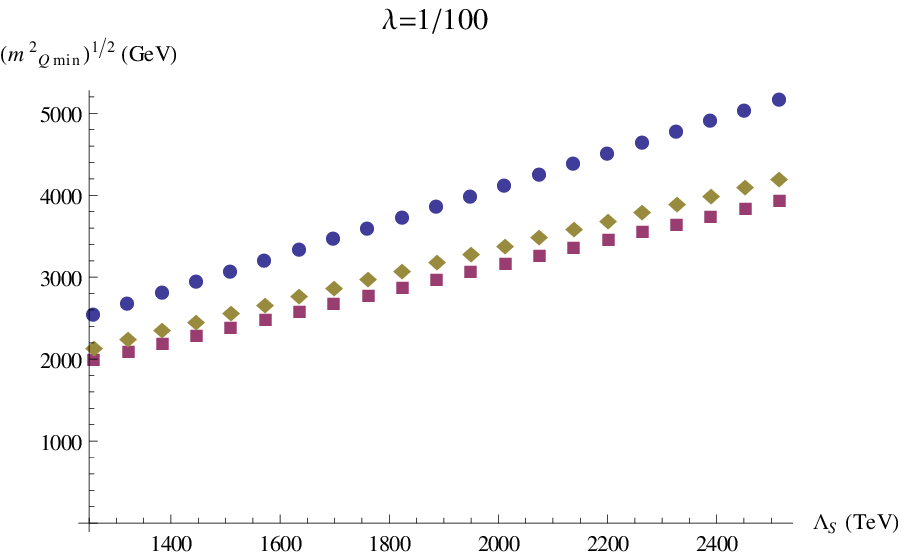}
\end{minipage}
\caption{\label{fig:tachyongm} Minimum value of the boundary value of $(m^2_Q)^{1/2}$ needed to avoid tachyonic soft masses at 500 GeV in terms of the high scale $\Lambda_S$, in minimal (left) and nonminimal (right) Effective SUSY scenarios with the boundary conditions of eq.~\eqref{eq:gmbc}, for $\lambda=1$ (upper plots) and $\lambda=1/100$ (lower plots). The upper blue dots correspond to the the MSSM  $ \overline{\rm DR}$ RG flow, the diamond-shaped marks represent the results with the flow implementing  decoupling and including threshold effects, while the boxes denote the results when using the flow implementing decoupling but ignoring threshold effects. The choices of $\Lambda_S$ correspond to heavy sparticles betwen 10 and 20 TeV.}
\end{figure}

Now, all these calculations did not take into account finite threshold effects from the heavy particles, which have been shown here to be large and predominantly negative, so that they will force an increase of the bounds and demand more fine-tuning in the theories. In the same spirit as in ref.~\cite{Tamarit:2012ry}, bounds can be obtained by considering boundary conditions inspired by msugra and gauge mediation but allowing for a large hierarchy between the masses of the sparticles of the first two generations and those of the third generation. The msugra-inspired boundary conditions, set at a SUSY breaking scale $\Lambda_S$, are
\begin{align}
\label{eq:msugrabc}
\begin{array}{c}
\text{ minimal Effective SUSY}\\
\mu=\tilde m_1=\tilde m_2=\tilde m_3=m_F,\\
 { m^2_{q/u/d/l/e}}_{11}= { m^2_{q/u/d/l/e}}_{22}= {m^2_{d/l/e}}_{33}=\Lambda^2,\\
{ m^2_{q/u}}_{33}=m_s^2,\\
\frac{a_u}{y_t}=\frac{a_d}{y_b}=\frac{a_l}{y_\tau}=a_0,
\end{array}\quad\begin{array}{c}
\text{ nonminimal Effective SUSY}\\
\mu=\tilde m_1=\tilde m_2=\tilde m_3=m_F,\\
 { m^2_{q/u/d/l/e}}_{11}= { m^2_{q/u/d/l/e}}_{22}=\Lambda^2,\\
{ m^2_{q/u/d/l/e}}_{33}=m_s^2,\\
\frac{a_u}{y_t}=\frac{a_d}{y_b}=\frac{a_l}{y_\tau}=a_0,
\end{array}
\end{align}
while the ones resembling gauge mediation, also set at a scale $\Lambda_S$, are
\begin{align}
\label{eq:gmbc}
\begin{array}{c}
 \text{minimal Effective SUSY}\\
\tilde m_i=g^2_i\Lambda_g,\\
{ m^2_{q/u/d/l/e}}_{11}={ m^2_{q/u/d/l/e}}_{22}={ m^2_{d/l/e}}_{33}=\lambda\frac{\Lambda_S^2}{16\pi^2},\\
{ m^2_{i}}_{33}= \Lambda^2_G\sum_k g_k^4 C^k_2(i),\,i=q,u,\\
a_u=a_d=a_l=0,
\end{array}\quad\begin{array}{c}
 \text{nonminimal Effective SUSY}\\
\tilde m_i=g^2_i\Lambda_g,\\
{ m^2_{q/u/d/l/e}}_{11}={ m^2_{q/u/d/l/e}}_{22}=\lambda\frac{\Lambda_S^2}{16\pi^2},\\
 { m^2_{i}}_{33}= \Lambda^2_G\sum_k g_k^4 C^k_2(i),\\
a_u=a_d=a_l=0,
\end{array}
\end{align}
Regarding the boundary conditions of eq.~\eqref{eq:msugrabc}, figure \ref{fig:tachyonmsugra} shows the resulting lower bound in the mass parameter $m_s$ in terms of the scale $\Lambda_S$ for minimal and nonminimal Effective SUSY scenarios, using the MSSM $ \overline{\rm DR}$ RG equations without threshold contributions, the decoupled RG flow of ref.~\cite{Tamarit:2012ie} without thresholds, and finally the decoupled RG flow together with the threshold contributions presented in this paper applied at the scale at which the heavy sparticles are integrated out. $m_F$ was fixed at 1 TeV, the heavy scalars at 20 TeV, and $a_0$ at 0. If the boundary conditions of eq.~\eqref{eq:gmbc} are used, figure \ref{fig:tachyongm} shows analogous results for the lower bound of the soft mass $m^2_Q$ at the scale $\Lambda_S$ with respect to this scale, for two different values of $\lambda$; $\Lambda_g$ was kept at 1 TeV, and the choices of $\Lambda_S$ correspond to heavy sparticles between 10 and 20 TeV. In the literature, and 
in spectrum calculators for SUSY scenarios, it is customary to compute physical sparticle masses at a scale near the stop mass in order to minimize theoretical errors; here for simplicity it was chosen to probe for tachyons at a scale of $500$ GeV.


The results show that the inclusion of the threshold effects in the decoupled RG analysis slightly increases the lower mass bounds obtained by demanding the absence of tachyons, but these bounds still remain well below the ones obtained with the MSSM $ \overline{\rm DR}$ RG flow without decoupling.

\begin{figure}[h]\centering
 \begin{minipage}{.5\textwidth}\centering
   \includegraphics[scale=.9]{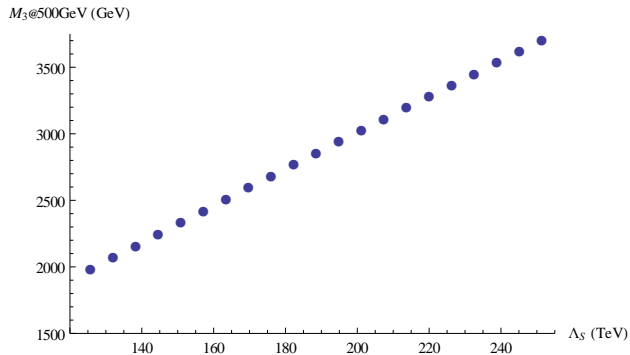}
\end{minipage}
\caption{\label{fig:tachyongaugino} Minimum value of the gluino mass $\tilde m_3$ at $500$ GeV needed to avoid tachyonic soft masses at the same scale in terms of the high scale $\Lambda_S$, in minimal Effective SUSY scenarios with the boundary conditions of eq.~\eqref{eq:gaugbc} for $\lambda=1$. }
\end{figure}

Another interesting set of boundary conditions concerns models with an Effective SUSY spectrum in which the light soft masses are generated through gaugino mediation, i.e., they arise through the RG effects of nonzero gaugino masses. Some examples can be found in refs.~\cite{Craig:2011yk,Craig:2012hc,Cohen:2012rm}, in which either deconstruction~\cite{Craig:2011yk,Craig:2012hc} or conformal sequestering \cite{Cohen:2012rm} are used to suppress light soft masses. The large, negative threshold effects that are the central subject of this paper may force unnatural fine-tuning in models of this type: integrating out the heavy sparticles produces large tachyonic contributions to the suppressed soft masses, which may not be compensated by the RG effects of gaugino masses unless these are unnaturally large. Again, one can obtain lower bounds for gaugino masses  by using simplified boundary conditions. In the spirit of gaugino mediation with heavy first and second generation scalars, one may 
consider the following ones at a scale $\Lambda_S$,
\begin{align}
\label{eq:gaugbc}
\begin{array}{c}
 \text{minimal Effective SUSY}\\
\tilde m_i=g^2_i\Lambda_g,\\
{ m^2_{q/u/d/l/e}}_{11}={ m^2_{q/u/d/l/e}}_{22}={ m^2_{d/l/e}}_{33}=\lambda \frac{\Lambda_S^2}{16\pi^2},\\
{ m^2_{i}}_{33}=0,\,i=q,u,\\
a_u=a_d=a_l=0,
\end{array}\quad\begin{array}{c}
 \text{nonminimal Effective SUSY}\\
\tilde m_i=g^2_i\Lambda_g,\\
{ m^2_{q/u/d/l/e}}_{11}={ m^2_{q/u/d/l/e}}_{22}=\lambda\frac{\Lambda_S^2}{16\pi^2},\\
 { m^2_{i}}_{33}= 0,\\
a_u=a_d=a_l=0.
\end{array}
\end{align}
The resulting minimum values of the gluino mass $\tilde m_3$ evaluated at 500 GeV are shown in figure \ref{fig:tachyongaugino} in terms of $\Lambda_S$ for minimal Effective SUSY scenarios, in the case $\lambda=1$. It is apparent that demanding no tachyonic charged/colored sparticles requires very heavy gluinos, at 2 TeV or heavier for heavy scalars at 10 TeV or above. Decreasing $\lambda$ implies raising the SUSY breaking scale for a fixed value of the heavy masses, which will only raise the bound on the gluino mass, as there will be more decades of MSSM RG running  driving the light soft masses towards negative values (see for example fig.~\ref{fig:tachyongm}). The case of nonminimal scenarios is rather hopeless; this time the mass running deeper into negative values is $m^2_L$ , and the bounds for $\tilde m_3$ reach 8 TeV and higher --eventually crossing the heavy particle threshold, so that the analysis would have to be modified. Alternatively, one may fix the heavy masses at a fixed value, for example at 
10 TeV with $\Lambda_g$ at 3 TeV, and probe $\lambda$ in order to obtain the maximum value allowed for the scale of SUSY breaking when demanding the absence of tachyonic masses at low scales; the resulting lower bound on $\Lambda_S$ is as low as 11 TeV. 

\section{Summary and conclusions\label{sec:models}}
  This paper presents results for finite threshold contributions to the soft masses of light scalars caused by loops involving heavy sparticles in Effective SUSY scenarios, and analyzes their influence in bounds for squarks and gauginos obtained by demanding the absence of tachyonic squarks and sleptons. In contrast with previous results in the literature, nonzero tree-level values for the soft masses of light squarks and gauginos were considered inside the two-loop diagrams contributing to the threshold corrections. It was shown that in the limit of degenerate heavy fields --possibly including a heavy Higgs mostly aligned with $H_d$--  the known one-loop corrections are mainly determined by the hypercharge coupling and small Yukawas and may be negative at the threshold scale of the heavy particles. In this limit in which the heavy fields couple to the light ones through small Yukawas, the two-loop diagrams are dominated by the contributions involving the gauge couplings, which were calculated ignoring 
mixing 
among the heavy states and considering nonzero tree-level masses for the light scalars and gauginos; the result is given in eqs.~\eqref{eq:2loopmassive} and \eqref{eq:2loophypercharge}.

  These two-loop contributions turn out to be quite significant, and they take negative values at the scale of the heavy fields, thus invalidating the na\"ive intuition that the heavy fields could act as ``messengers'' of SUSY breaking for the light scalars. In nonminimal scenarios,  for the soft mass of the left-handed third generation squark doublet, they range from around $-250^2\, {\rm GeV}^2$ to $-480^2 {\rm GeV}^2$ for heavy sparticles between 5 and 10 TeV, if light scalars and gauginos are kept at around $500$ GeV (see figure \ref{fig:2loopthresh})--  and thus should not be ignored when analyzing the properties of Effective SUSY models. Fig.~\ref{fig:2loopthresh} also shows that gaugino masses have a sizable impact in the threshold corrections, which become more negative for larger gaugino masses --a gluino mass of around 1 TeV enhances the threshold correction by 20\% or more. The dependence on the light scalar masses is weaker.
  
The large negative threshold corrections to the light soft masses add to the already known negative 2 loop effects in the RG flow due to the heavy sparticles, which endanger the stability of the electroweak vacuum and may give rise to charged or colored vacua, the avoidance of which forces a lower bound on the light soft masses at the SUSY breaking scale, which translates into a lower bound in fine-tuning. Tachyon bounds for squark masses were reanalyzed, taking into account the threshold effects and using an RG flow implementing decoupling. It was shown that the bounds are slightly increased, but the use of the decoupled RG flow  still guarantees that the former remain lower than the ones obtained by using the MSSM $\overline{\rm DR}$ RG equations without integrating out heavy sparticles. This strengthens the case for the need of implementing decoupling in precision calculations in models with hierarchical spectra.

In the case of models in which the light soft masses arise from gaugino mediation, and are thus approximately zero at the SUSY breaking scale, by demanding again the absence of tachyonic squarks/sleptons one may obtain lower bounds for gaugino masses. It was shown that for simple boundary conditions in minimal Effective SUSY scenarios (assuming for example that the SUSY breaking scale is related the scale of the heavy masses by a loop factor), these bounds require gluinos above 2 TeV for heavy squarks at 10 TeV or higher. In nonminimal scenarios the bound is rather more stringent, requiring gluinos above 8 TeV. Alternatively, if the SUSY scale is left to vary with heavy fields fixed at 10 TeV and the boundary value for $\tilde m_3$ fixed at 3 TeV, then the said scale has to be barely above the mass of the heavy fields. These constraints may be avoided in deconstructed SUSY breaking models in which the heavy and light scalars are charged under different gauge groups, as in refs. \cite{Craig:2011yk,Craig:
2012hc},
 if the scale at which these groups are higgsed to the diagonal is below the mass of the heavy sparticles. This would imply the presence of new fields beyond the MSSM under the scale of the heavy sparticles, which would alter the RG flow.

Finally, it should be commented that the negative threshold contributions also affect the light Higgs fields, so that they might play an important role in the breaking of electroweak symmetry breaking.
\section*{Acknowledgements}
The author wishes to thank the members of the Particle Physics group at Perimeter Institute for useful conversations. Research
at the Perimeter Institute is supported in part by the Government of Canada through
NSERC and by the Province of Ontario through MEDT. This work was financed in part by the Spanish Ministry of Science and Innovation through project FPA2011-24568.
\appendix
\section{Some two loop integrals in dimensional regularization\label{app:integrals}}

The calculations were performed in dimensional regularization with $d=4-2\epsilon$ dimensions. Using standard manipulations, all integrals can be written in terms of one loop integrals and two-loop ones involving three propagators. If at least one of the propagators is massless, the integrals can be obtained from the following formula, obtained by applying Mellin-Barnes techniques (see for example ref.~\cite{Smirnov:2006ry}):
\begin{align}
 \nonumber&I[m_1,m_2;n_1,n_2,n_3]\equiv\int \frac{d^dpd^dq}{(2\pi)^{2d}}\frac{1}{(p^2-m_1^2)^{n_1}(q^2-m_2^2)^{n_2}((p+q)^2)^{n_3}}=\\
 \nonumber&\frac{(-1)^{(1-d)}2^{-2d}\pi^{-d}(-m_1^2)^{d-n_1-n_2-n_3}\Gamma[\frac{d}{2}-n_3]\Gamma[a]\Gamma[b]\Gamma[-\frac{d}{2}+n_1+n_3]}{\Gamma\left[\frac{d}{2}\right]\Gamma[n_1]\Gamma[n_2]\Gamma[c]}\,\, {}_2F_1\Big[a,b,c,1-\frac{m^2_2}{m_1^2}\Big],\\
 &a=-\frac{d}{2}+n_2+n_3,\,\,\,b=-d+n_1+n_2+n_3,\,\,\,c=-d+n_1+n_2+2n_3, \label{eq:masterI}
\end{align}
where  ${}_2F_1$ is a hypergeometric function in the usual notation. Relevant cases are, taking massless or degenerate limits when necessary,

\begin{align*}
 &I[m_1,m_2;1,1,1]=\\
 &\frac{1}{\Gamma[2-\epsilon]}16^{-2+\epsilon} \left(m_1 m_2\right)^{-2 \epsilon} \pi ^{-4+2 \epsilon} \Gamma[1-\epsilon] \left\{{m_2}^2 \Gamma[-1+\epsilon] \Gamma[\epsilon]+\left(\frac{{m_2}}{{m_1}}\right)^{2 \epsilon} \left(-{m_2}^2\right.\right.\\
 &+\left({m_1}^2+(-1+2 \epsilon) {m_2}^2\right) \pi  \csc[\epsilon \pi ] \Gamma[-1+2 \epsilon]+2 \log\frac{{m_2}}{{m_1}} \left({m_2}^2+({m_1}-{m_2}) ({m_1}+{m_2}) \times\right.\\
 &\left.\left.\left.\times\log\left[1-\frac{{m_2}^2}{{m_1}^2}\right]\right)+({m_1}-{m_2}) ({m_1}+{m_2}) \text{Li}_2\left[\frac{{m_2}^2}{{m_1}^2}\right]\right)\right\}+O(\epsilon),\\
& I[m,m;1,1,1]=\frac{16^{-2+\epsilon} m^{2-4 \epsilon} \pi ^{-3+2 \epsilon} \csc[\epsilon \pi ] \Gamma[\epsilon]}{(-1+2 \epsilon) \Gamma[2-\epsilon]},\\
 &I[m_1,m_2;1,2,1]=\frac{{m_1}^{-4\epsilon} (4 \pi )^{-4+2\epsilon} \Gamma\left[1-\epsilon\right]}{\Gamma[2-\epsilon]}\left\{\frac{\pi ^2}{3}\!-\!\left(\frac{{m_2}^2}{{m_1}^2}\right)^{\!\!-2 \epsilon} \pi  \csc[\epsilon \pi ] \Gamma[2 \epsilon]+\frac{1}{2} \log\frac{{m_2}^2}{{m_1}^2}\times\right.\\
 &\left.\times\left(\log\frac{{m_2}^2}{{m_1}^2}-2 \log\left[1-\frac{{m_2}^2}{{m_1}^2}\right]\right)-\text{Li}_2\left[\frac{{m_2}^2}{{m_1}^2}\right]\right\}+O(\epsilon),\\
 &I[m,m;1,1,2]=\frac{2^{-9+4 \epsilon}  {m}^{-4 \epsilon} \pi ^{-3+2 \epsilon} \csc[\epsilon \pi ] \Gamma[\epsilon]}{(1+2 \epsilon) \Gamma[2-\epsilon]},\\
 &I[m,0;1,1,1]=-2^{-7+4 \epsilon} {m}^{2-4 \epsilon} \pi ^{-3+2 \epsilon} \csc[\epsilon \pi ] \Gamma[-2+2 \epsilon],\\
\end{align*}
where $\rm Li_2$ is the usual dilogarithm function. It should be noticed that the finite part of $I[m_1,m_2,1,1,2]$ that results from the above formulae is different from the $m_\epsilon$-independent contribution to the finite part of the corresponding result in ref.~\cite{Martin:1996zb}, in which $m_\epsilon$ was introduced as an explicit infrared regulator for the zero mass propagators. This latter result was used in the calculations of ref.~\cite{Hisano:2000wy}, which explains the discrepancy of their result with formula \eqref{eq:2loopmassless}, which arises from the different choices of regularization.

The following integral with three massive propagators is also relevant for the calculation \cite{Davydychev:1992mt}:
\begin{align*}
 &
\int \frac{d^dpd^dq}{(2\pi)^{2d}}\frac{1}{(p^2-m_1^2)(q^2-m_1^2)((p+q)^2-m_2^2)}=\\
&\frac{2^{-8+4 \epsilon}  {m_1}^{2-4 \epsilon} \pi ^{-4+2 \epsilon} \Gamma[1+\epsilon]^2}{(1 - 2 \epsilon) (1 - \epsilon)}\left(-\frac{1+2z}{\epsilon^2}+\frac{4z\log(4z)}{\epsilon }-2z\log^2(4z)+2 \left(1-z\right) \Phi\left[z\right]\right)+O(\epsilon),\\
&z=\frac{m_2^2}{4m_1^2}.
\end{align*}
For $z<1$ one may write
\begin{align}
 \phi[z]=4\sqrt{\frac{z}{1-z}}\,{\rm Cl_2}(2\arcsin\sqrt{z}),\,\,\,{\rm Cl_2}(z)=-\int_0^z dt \log\left|2\sin\frac{t}{2}\right|.
 \label{eq:phidef}
\end{align} 
\newpage 
\bibliographystyle{h-physrev}
\bibliography{ESthresholdsref}

\end{document}